\documentclass[twocolumn,pra,aps,showpacs]{revtex4}
\usepackage{}
\usepackage{amsmath}
\usepackage{color}%
\usepackage{graphicx}

\begin{document}

\title{Photon scattering by an atomic ensemble coupled to a one-dimensional nanophotonic waveguide}

\author{Guo-Zhu Song$^{1,2,3}$, Ewan Munro$^{1}$, Wei Nie$^{1}$, Fu-Guo Deng$^{2,3}$, Guo-Jian Yang$^{3}$, and Leong-Chuan Kwek$^{1,4,5,6}$\footnote{Corresponding author: cqtklc@nus.edu.sg} }

\address{$^{1}$Centre for Quantum Technologies, National University of Singapore, 3 Science Drive 2, Singapore 117543\\
$^{2}$NAAM-Research Group, Department of Mathematics, Faculty of Science, King Abdulaziz University, P.O. Box 80203, Jeddah 21589, Saudi Arabia\\
$^{3}$Department of Physics, Applied Optics Beijing Area Major Laboratory, Beijing Normal University, Beijing 100875, China\\
$^{4}$Institute of Advanced Studies, Nanyang Technological University, Singapore 639673\\
$^{5}$National Institute of Education, Nanyang Technological University, Singapore 637616\\
$^{6}$MajuLab, CNRS-UNS-NUS-NTU International Joint Research Unit, UMI 3654, Singapore}

\date{\today }

\begin{abstract}
We theoretically investigate the quantum scattering of a weak coherent input field interacting with an ensemble of $\Lambda$-type three-level atoms coupled to a one-dimensional waveguide. With an effective non-Hermitian Hamiltonian, we study the collective interaction between the atoms mediated by the waveguide mode.
In our scheme, the atoms are randomly placed in the lattice along the axis of the one-dimensional waveguide. Many interesting optical properties occur in our waveguide-atom system, such as electromagnetically induced transparency (EIT). We quantify the influence of decoherence originating from both dephasing and population relaxation, and analyze the effect of the inhomogeneous broadening on the transport properties of the incident field. Moreover, we observe that strong photon-photon correlation with quantum beats can be generated in the off-resonant case, which provides an effective method for producing non-classical light in experiment. With remarkable progress in waveguide-emitter system, our scheme may be experimentally feasible in the near future.
\end{abstract}

\maketitle


\section{Introduction}

Since single photons have long coherence times, they are considered as good candidates for quantum information processing \cite{Kimble2008} and quantum memory \cite{Ohlsson2003,Nilsson2005}. On the other hand, atoms are chosen as stationary qubits due to their potential scalability and stability. In the past decades, strong photon-atom interaction has been achieved by confining the photons in the high-quality optical microcavity \cite{Mabuchi,Walther}. Recently,  photon transport in a one-dimensional (1D) waveguide coupled to quantum emitters, known as waveguide quantum electrodynamics (QED), has been widely studied \cite{ShenOL2005,ShenPRL2005,ShenPRA2007,AkimovNature2007,ShenPRL2007,Faraon2007,TsoiPRA2008,ZhouPRL2008,TsoiPRA2009,BajcsyPRL2009,LongoPRL2010,WitthautNJP2010,AstafievScience2010,Abdumalikov2010,
BabinecNat2010,AngelakisPRL2011,ZhengPRL2011,RoyPRL2011,BleusePRL2011,BradfordPRL2012,PletyukhovNJP2012,AngelakisPRL2013,DayanScience2013,ZhengBarangerPRL2013,RoySR2337,TudelaNAT2015,GreenbergPRA2015,YCH2015,EWAN,Roy2017RMP}, which provides a promising candidate for realizing strong light-matter interactions. This 1D waveguide can be implemented by surface plasmon nanowire \cite{AkimovNature2007}, optical nanofibers \cite{AngelakisPRL2011,AngelakisPRL2013,DayanScience2013}, superconducting microwave transmission lines \cite{WallraffNature2004,Abdumalikov2010,AstafievScience2010,Hoi2011,HoiPRL2012,LooSci2013}, photonic crystal waveguide \cite{Faraon2007,EWAN,Frandsen2006}, and diamond waveguide \cite{BabinecNat2010,ClaudonPhoton2010}.

Using real-space description of the Hamiltonian and the Bethe-ansatz method, Shen and Fan studied the transport properties of a single photon and two photons scattered by an emitter embedded in a 1D waveguide \cite{ShenOL2005,ShenPRL2005,ShenPRA2007}. Interestingly, due to destructive quantum interference,
a photon with frequency resonant to the two-level quantum emitter can be completely reflected when the free-space emission is not considered.
Later, several approaches were proposed to calculate single-photon transport in a 1D waveguide coupled to a  two-level emitter, such as the input-output theory \cite{FanPRA2010}, Lippmann-Schwinger scattering method \cite{HuangPRA2013}, and the time-dependent theory \cite{ChenNJP2011}. Moreover, the scattering of a single photon by a driven $\Lambda$-type three-level emitter coupled to a 1D waveguide has been also studied \cite{RoyPRL2011,WitthautNJP2010,DRoyPRA2014}. In contrast to the single emitter case, a single photon scattered by multiple emitters can give rise to much richer behavior due to interference effects from multiple scattering. By solving the eigenvectors of the Hamiltonian in the single excitation subspace, Tsoi and Law \cite{TsoiPRA2008} investigated the interaction between a single photon and a finite chain of $N$ equally spaced two-level atoms inside a 1D waveguide. Compared with the single emitter case, they found that the transmission spectrum can be strongly modified in the collective many-body system, and the positions of the transmission peaks are
determined by the spacing between neighboring atoms. Later, Liao \emph{et al.} \cite{LZyangPRA2015} studied this system with a time-dependent theory, where many interesting phenomena occur such as Fano-like interference, superradiant effects, and photonic band-gap effects. In 2012, Chang \emph{et al.} \cite{Chang2012} demonstrated that two sets of equally spaced atomic chains coupled to a tapered nanofiber can form an effective cavity, which has long relaxation time and is highly dispersive compared to a conventional cavity.

Motivated by the important works mentioned above, we focus on the scattering property  of a weak coherent input field interacting with an ensemble of $\Lambda$-type three-level atoms coupled to a 1D waveguide. Different from the previous work where the emitters are equally spaced, the atoms are randomly located in the lattice along the axis of the
1D waveguide in our system, which closely corresponds to the experimental condition that the positions of atoms can not be manipulated precisely due to inevitable technological spreading of the parameters. Since the transmission and reflection spectra are fluctuant with the changeable configurations of the atomic positions and single-shot spectrum
is often unavailable due to finite trap lifetimes, we take the average values from a large sample of atomic spatial distributions and calculate the statistical properties of the system.

In this paper, we first assume that the input field is monochromatic and calculate the transport properties of a three-level atomic ensemble coupled to a 1D waveguide. We analyze the effect of decoherence originating from both population relaxation and dephasing, and quantify the influence of the inhomogeneous broadening on the transmission and reflection spectra of the incident field. Then, we consider a photon pulse with Gaussian shape and study the optical properties with the parameters of our system, such as the Rabi frequency of the driving field, the coupling strength between atomic ensemble and the 1D waveguide, lattice constant, and the number of atoms.  Besides, since atoms are randomly placed in the lattice, we analyze the variance of the transmission as a function of the frequency detuning, concluding that the influence of atomic spatial distributions on transport properties changes with frequency detuning. Finally, we calculate the second-order correlation function in off-resonant case, and observe non-classical behavior in our system. We find that, with strong driving field, both anti-bunching and bunching appear in the transmitted field, while only  bunching occurs in the reflected field. Moreover, quantum beats (oscillations) \cite{ZhengPRL2013} emerge in the photon-photon correlation function of the reflected and transmitted fields. In fact, our system provides an effective method for producing non-classical light in experiment.

The paper is organized as follows: In Sec. \ref{MODEL}, we give the model and present the derivation of the effective Hamiltonian for the system composed of an ensemble of three-level atoms and the propagating field in a 1D waveguide. In Sec. \ref{RESULTS}, we study the transport properties of a weak coherent input field with the influence of decoherence and inhomogeneous broadening, the variance caused by atomic spatial distributions, and photon-photon correlation in the off-resonant case. Finally, a summary is shown in Sec. \ref{discussion}.

\section{MODEL AND HAMILTONIAN}  \label{MODEL}

\begin{figure}
\centering\includegraphics[width=7.5cm]{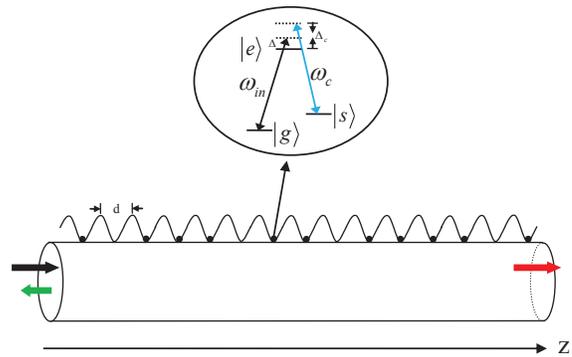}
\caption{Schematic diagram for the scattering of an input field off an ensemble of $\Lambda$-type three-level atoms (black dots) coupled to a 1D waveguide (the cylinder). A weak coherent field (black arrow) is input from left to interact with the atomic ensemble, which generates a transmitted part (red arrow) and a reflected part (green arrow). The wavy line denotes the lattice, and $d$ is the lattice constant. Each trap site is either empty or contains a single atom \cite{HLsorensen2016}.  } \label{figure1}
\end{figure}

In this section, we consider a system composed of an ensemble of $\Lambda$-type three-level atoms randomly located in a lattice of period $d$ along the waveguide, as shown in Fig. \ref{figure1}. We assume that the transition with the resonance frequency $\omega_{a}$ between ground state $|g\rangle$ and excited state $|e\rangle$ is coupled to the mode of the 1D waveguide, and the transition $|e\rangle\leftrightarrow|s\rangle$ is driven by a classical  field with the Rabi frequency $\Omega_{c}$. The Hamiltonian of the full system with the rotating-wave approximation in real space reads (taking $\hbar=1$) \cite{ShenOL2005}
\begin{eqnarray}     \label{eqa1}       
\begin{split}
H=\;&ic\!\int \!dz\big[a_{_{L}}^{\dag}(z)\frac{\partial a_{_{L}}(z)}{\partial z} - a_{_{R}}^{\dag}(z)\frac{\partial a_{_{R}}(z)}{\partial z}\big]\\
&+ \sum\limits_{j}^n\big[\omega_{a}\sigma_{ee}^{j}\!+\!\omega_{s}\sigma_{ss}^{j}\!-\!\Omega_{c}(\sigma_{es}^{j}e^{-i\omega_{c} t}+h.c.)\big]\\
&-\tilde{g}\!\!\int \!\!dz\sum\limits_{j}^n\delta(z-z_{j})\big\{\sigma_{eg}^{j}[a_{_{R}}(z)+a_{_{L}}(z)]+h.c.\big\},
\end{split}
\end{eqnarray}
where $a_{_{R}}$ ($a_{_{L}}$) denotes the annihilation operator of right (left) propagating field, and $\tilde{g}=\sqrt{2\pi}g$. $g$ is the coupling strength between the atom and the waveguide mode, assumed to be identical for all the atoms. $\omega_{c}$ is the frequency of the driving field. Here, we take the energy of the ground state $|g\rangle$ to be zero, and $\omega_{s}$ is the energy of the level $|s\rangle$. The atomic operator $\sigma_{\alpha\beta}^{j}=|\alpha_{j}\rangle\langle\beta_{j}|$ with $\alpha,\beta=g,e,s$ being the energy eigenstates of the $j^{th}$ atom.

By calculating the commutators with $H$, we can obtain the Heisenberg equations of the motion for the atomic operators
\begin{eqnarray}     \label{eqa2}       
\dot{\sigma}_{ge}^{j}\!\!\!&=&\!\!\!-i\omega_{a}\sigma_{ge}^{j}\!\!+\!i\Omega_{c}\sigma_{gs}^{j}e^{-i\omega_{c} t}\!\!+\!i\tilde{g}(\sigma_{gg}^{j}\!\!-\!\!\sigma_{ee}^{j})[a_{_{R}}(z_{j})\!+\!a_{_{L}}(z_{j})],\nonumber\\
\dot{\sigma}_{gs}^{j}\!\!\!&=&\!\!\!-i\omega_{s}\sigma_{gs}^{j}+i\Omega_{c}\sigma_{ge}^{j}e^{i\omega_{c} t}-i\tilde{g}\sigma_{es}^{j}[a_{_{R}}(z_{j})\!+\!a_{_{L}}(z_{j})],\nonumber\\
\dot{\sigma}_{es}^{j}\!\!&=&\!\!i\omega_{c}\sigma_{es}^{j}\!\!+\!i\Omega_{c}(\sigma_{ee}^{j}\!-\!\sigma_{ss}^{j})e^{i\delta t}\!-\!i\tilde{g}\sigma_{gs}^{j}[a_{_{R}}^{\dagger}(z_{j})\!+\!a_{_{L}}^{\dagger}(z_{j})].
\end{eqnarray}
Using the same method, we can also write the Heisenberg equations of the motion for the photons
\begin{eqnarray}     \label{eqa3}       
\begin{split}
(\frac{1}{c}\frac{\partial}{\partial t}+\frac{\partial}{\partial z})a_{_{R}}(z)=\frac{i\tilde{g}}{c}\sum\limits_{j}^n\delta(z-z_{j})\sigma_{ge}^{j},\\
(\frac{1}{c}\frac{\partial}{\partial t}-\frac{\partial}{\partial z})a_{_{L}}(z)=\frac{i\tilde{g}}{c}\sum\limits_{j}^n\delta(z-z_{j})\sigma_{ge}^{j},
\end{split}
\end{eqnarray}
where $c$ is the velocity of the traveling photon. Then, the Heisenberg equations for $a_{_{R}}$ ($a_{_{L}}$) can be integrated, and we get the real-space wave equation
\begin{eqnarray}     \label{eqa4}       
\begin{split}
a_{_{R}}(z,t)\!=\!a_{_{R,in}}(z\!-\!ct)\!+\!\frac{i\tilde{g}}{c}\!\sum\limits_{j}\!\theta(z\!-\!z_{j})\sigma_{ge}^{j}(t\!-\!\frac{z\!-\!z_{j}}{c}),\\
a_{_{L}}(z,t)\!=\!a_{_{L,in}}(z\!+\!ct)\!+\!\frac{i\tilde{g}}{c}\!\sum\limits_{j}\!\theta(z_{j}\!-\!z)\sigma_{ge}^{j}(t\!-\!\frac{z_{j}\!-\!z}{c}).
\end{split}
\end{eqnarray}
Here, the first term $a_{_{R,in}}$ ($a_{_{L,in}}$) represents the freely traveling field in the waveguide, while the second term corresponds to the contribution of the field emitted by the atomic ensemble. $\theta(z)$ is the Heaviside step function. Since we are more interested in the scattered field induced by atoms, here we set $a_{_{R,in}}\!=\!a_{_{L,in}}\!=\!0$.
Inserting the above field equation into the Eq. (\ref{eqa2}), we get the equations for the atoms alone
\begin{eqnarray}     \label{eqa5}       
\begin{split}
\dot{\sigma}_{ge}^{j}=&-i\omega_{a}\sigma_{ge}^{j}+i\Omega_{c}\sigma_{gs}^{j}e^{-i\omega_{c} t}\\
&-\frac{\tilde{g}^{2}}{c}(\sigma_{gg}^{j}-\sigma_{ee}^{j})\sum\limits_{j,k}\sigma_{ge}^{k}(t-\frac{|z_{j}-z_{k}|}{c}),\\
\dot{\sigma}_{gs}^{j}=&-i\omega_{s}\sigma_{gs}^{j}+i\Omega_{c}\sigma_{ge}^{j}e^{i\omega_{c} t}\\
&+\frac{\tilde{g}^{2}}{c}\sigma_{es}^{j}\sum\limits_{j,k}\sigma_{ge}^{k}(t-\frac{|z_{j}-z_{k}|}{c}),\\
\dot{\sigma}_{es}^{j}=&\;\;i\omega_{c}\sigma_{es}^{j}+i\Omega_{c}(\sigma_{ee}^{j}-\sigma_{ss}^{j})e^{i\omega_{c} t}\\
&-\frac{\tilde{g}^{2}}{c}\sigma_{gs}^{j}\sum\limits_{j,k}\sigma_{eg}^{k}(t-\frac{|z_{j}-z_{k}|}{c}).
\end{split}
\end{eqnarray}
Then, by transforming to the slow-varying frame, we can define the three following quantities:
\begin{eqnarray}     \label{eqa6}       
\begin{split}
\sigma_{ge}^{j}(t)&=S_{ge}^{j}(t)e^{-i\omega_{in}t},\;\;\sigma_{gs}^{j}(t)=S_{gs}^{j}(t)e^{-i(\omega_{in}-\omega_{c})t},\\
\sigma_{es}^{j}(t)&=S_{es}^{j}(t)e^{i\omega_{c} t},
\end{split}
\end{eqnarray}
where $\omega_{in}$ is the frequency of the incident field.

When the atomic resonance frequency $\omega_{a}$ is far away from the cutoff frequency of the waveguided mode,
and the photon has a narrow bandwidth in vicinity of $\omega_{a}$, we can adopt the linear dispersion approximation \cite{ShenPRA2009}.
Using this condition,  Eq. (\ref{eqa5}) is rewritten as
\begin{eqnarray}     \label{eqa7}       
\begin{split}
\dot{S}_{ge}^{j}=\;&i\Delta S_{ge}^{j}+i\Omega_{c}S_{gs}^{j}\\
&-\frac{\Gamma_{_{1D}}}{2}(S_{gg}^{j}-S_{ee}^{j})\sum\limits_{j,k}S_{ge}^{k}(t)e^{ik_{in}|z_{j}-z_{k}|},\\
\dot{S}_{gs}^{j}=\;&i(\Delta-\Delta_{c})S_{gs}^{j}+i\Omega_{c}S_{ge}^{j}\\
&+\frac{\Gamma_{_{1D}}}{2}S_{es}^{j}\sum\limits_{j,k}S_{ge}^{k}(t)e^{ik_{in}|z_{j}-z_{k}|},\\
\dot{S}_{es}^{j}=\;&i\Omega_{c}(S_{ee}^{j}-S_{ss}^{j})-\frac{\Gamma_{_{1D}}}{2}S_{gs}^{j}\sum\limits_{j,k}S_{eg}^{k}(t)e^{ik_{in}|z_{j}-z_{k}|},
\end{split}
\end{eqnarray}
where $\Delta\!\!=\!\!\omega_{in}-\omega_{a}$, and $\Gamma_{_{1D}}\!\!=\!\!2\tilde{g}^{2}/c$. $\Delta_{c}\!\!=\!\!\omega_{c}-\omega_{es}$ is the frequency detuning between the driving field and the transition $|s\rangle\!\leftrightarrow\!|e\rangle$. From the above equations, after eliminating the fields, we can get an effective Hamiltonian for the system
\begin{eqnarray}     \label{eqa8}       
H_{_{eff}}\!=\!\!\!\!&&-{{\sum\limits_{j}^n}}\big[\Delta S_{ee}^{j}\!+\!(\Delta-\Delta_{c})S_{ss}^{j}\big]\!-\!\Omega_{c}{{\sum\limits_{j}^n}}[(S_{es}^{j}\!+\!S_{se}^{j})]\nonumber\\
&&-i\frac{\Gamma_{_{1D}}}{2}{{\sum\limits_{j,k}^n}}e^{ik_{in}|z_{j}-z_{k}|}S_{eg}^{j}S_{ge}^{k}.
\end{eqnarray}

In the spirit of the quantum jump, spontaneous emission into free space other than the waveguide can be modeled by attributing an imaginary part $-i\Gamma'/2$ to the energies of the states of the atoms \cite{Carmichael1993}. Therefore, the system composed of the atomic ensemble and the 1D waveguide can be described by an effective non-Hermitian Hamiltonian
\begin{eqnarray}     \label{eqa9}       
\begin{split}
H_{non}\!=\!&-{{\sum\limits_{j}^n}}\big[(\Delta\!+\!i\Gamma_{e}^{'}/2)S_{ee}^{j}\!+\!(\Delta-\Delta_{c}) S_{ss}^{j}\\
&+\Omega_{c}(S_{es}^{j}+h.c.)\big]-i\frac{\Gamma_{_{1D}}}{2}{{\sum\limits_{j,k}^n}}e^{ik_{in}|z_{j}-z_{k}|}S_{eg}^{j}S_{ge}^{k},
\end{split}
\end{eqnarray}
where $\Gamma_{e}^{'}$ is the decay rate of the state $|e\rangle$ into the free space, and $z_{j}$ is the position of the $j^{th}$ atom.

Here, we focus mainly on the propagation of a constant weak coherent probe field. The corresponding driving is given by $H_{dri}\!=\!\sqrt{\frac{c\Gamma_{_{1D}}}{2}}\mathcal {E}{{\sum\limits_{j}^n}}(S_{eg}^{j}e^{ik_{in}z_{j}}+S_{ge}^{j}e^{-ik_{in}z_{j}})$, where $\sqrt{\frac{c\Gamma_{_{1D}}}{2}}\mathcal {E}$ is the amplitude of the constant input field \cite{CanevaNJP2015}. Therefore, the whole system can be described by the total Hamiltonian $H\!=\!H_{non}+H_{dri}$. For a sufficiently weak input field ($\sqrt{\frac{c\Gamma_{_{1D}}}{2}}\mathcal {E}\!\ll\!\Gamma_{e}^{'}$), quantum jumps can be ignored \cite{EWAN}. Provided that all atoms are in the ground state $|g\rangle$ and a weak coherent field with the wavevector $k_{in}$ is incident from the left, with the input-output methods \cite{CanevaNJP2015}, we can obtain the transmitted ($T$) and reflected ($R$) fields
\begin{eqnarray}     \label{eqa10}       
\begin{split}
a_{_{out,T}}(z) =&\; \mathcal {E}e^{ik_{in}z}+i\sqrt{\frac{\Gamma_{_{1D}}}{2c}}{{\sum\limits_{j}^n}}S_{ge}^{j}e^{ik_{in}(z-z_{j})},\\
a_{_{out,R}}(z) =&\; i\sqrt{\frac{\Gamma_{_{1D}}}{2c}}{{\sum\limits_{j}^n}}S_{ge}^{j}e^{-ik_{in}(z-z_{j})},
\end{split}
\end{eqnarray}
where the transmitted (reflected) field is defined for $z>z_{_{R}}\equiv max[z_{i}]$ ($z<z_{_{L}}\equiv min[z_{i}]$). In fact,  the optical properties of the output field are determined  by the input field and the dynamics of the atom-waveguide system alone. Therefore, the reflection of the incident field for the steady state is calculated by
\begin{eqnarray}     \label{eqa11}       
\begin{split}
R=\!\frac{\langle\psi|a_{_{out,R}}^{\dagger}a_{_{out,R}}|\psi\rangle}{\mathcal {E}^{2}},
\end{split}
\end{eqnarray}
where $|\psi\rangle$ is the steady-state wavevector. For the transmitted field, the  equation is similar.

\section{RESULTS} \label{RESULTS}

\subsection{The transmission and reflection of the input field} \label{TR}

\begin{figure}
\centering\includegraphics[width= 8.6cm]{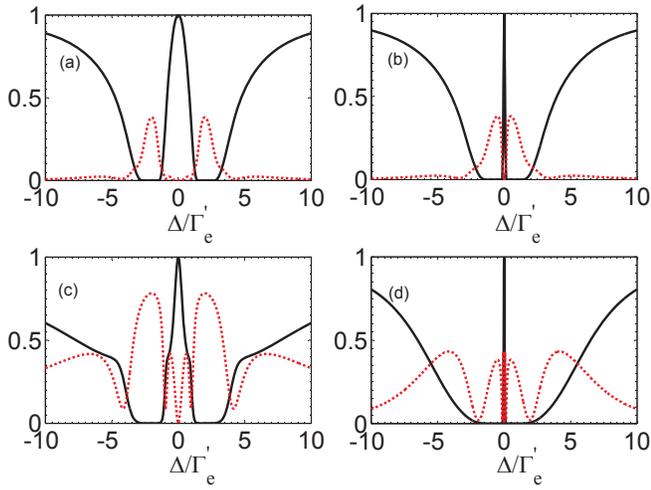}
\caption{The transmission $T$ (black solid line) and reflection $R$ (red dotted line) spectra of the input field as a function of the frequency detuning $\Delta/\Gamma_{e}'$ in two cases. Here, one case is that $n\!=\!10$ atoms are placed in a lattice of $N\!=\!10$ sites, i.e., 10 atoms are equally spaced with (a) $\Omega_{c}\!=\!2\Gamma_{e}'$, (b) $\Omega_{c}\!=\!0.5\Gamma_{e}'$. The other case is that $n=10$ atoms are randomly placed in a lattice of $N\!=\!200$ sites with  (c) $\Omega_{c}\!=\!2\Gamma_{e}'$, (d) $\Omega_{c}\!=\!0.5\Gamma_{e}'$. Parameters: (a)-(d) $\mathcal {E}=0.0001\sqrt{\frac{\Gamma_{_{1D}}}{2c}}$, $\Gamma_{_{1D}}\!=\!2\Gamma_{e}'$, $k_{in}d\!=\!\pi/2$, and $\Delta_{c}\!=\!0$.} \label{figure2}
\end{figure}

The quantum interference of a single-photon scattering with a chain of atoms inside a 1D waveguide has been studied in the previous works \cite{TsoiPRA2008,Leung2012,ChengPRA2017,LZyangPRA2015,Ruostekoski2017PRA}. In their calculations, the atoms are equally spaced with a deterministic separation $d$, which can be solved by the Bethe-ansatz approach \cite{TsoiPRA2008,ChengPRA2017}, the transfer matrix method \cite{Leung2012,Ruostekoski2017PRA}, and time-dependent dynamical theory \cite{LZyangPRA2015}. In this section, assuming that the input field is monochromatic, we study the scattering spectrum for $n\!=\!10$ three-level atoms randomly placed in a lattice of $N\!=\!200$ sites.
For comparison, we first give the transmission and reflection of the input field traveling through 10 equally spaced three-level atoms, as shown in Figs. \ref{figure2}(a)-(b). While, when 10 atoms are randomly placed in a lattice of $N\!\!=\!\!200$ sites, the results are quite different, and the calculations for one possible configuration of atomic positions are shown in Figs. \ref{figure2}(c)-\ref{figure2}(d). Compared with the first row of Fig. \ref{figure2}, the reflection spectrum of the input field is modified remarkably in the latter case, and more peaks may appear in some  specific configurations of atomic positions. In fact, the scattering property of the input field is influenced by atomic spatial distributions.

\begin{figure}
\centering\includegraphics[width= 8.65cm]{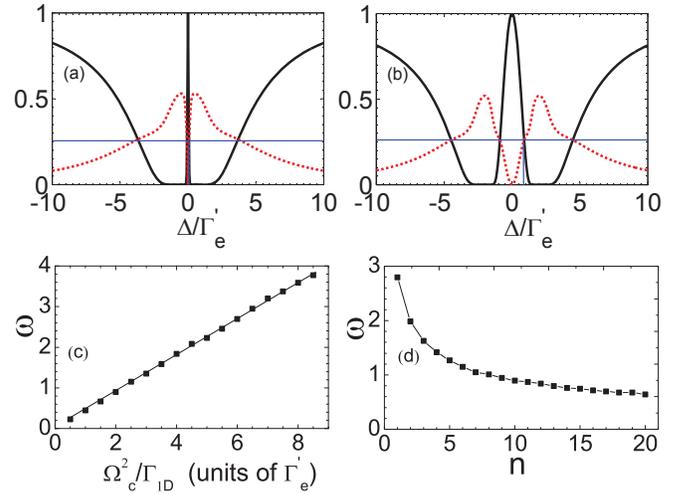}
\caption{The average transmission $\overline{T}$ (black solid line) and reflection $\overline{R}$ (red dotted line) spectra of the input field as a function of the frequency detuning $\Delta/\Gamma_{e}'$ for (a) $\Omega_{c}\!\!=\!\!0.5\Gamma_{e}'$, (b) $\Omega_{c}\!=\!2\Gamma_{e}'$. The width $w$ of the EIT window as a function of the parameter (c) $\Omega_{c}^2/\Gamma_{_{1D}}$, and the number of atoms (d) $n$. (a)-(c) $n\!\!=\!\!10$ atoms are randomly placed in a lattice of $N\!\!=\!\!200$ sites, (d) $N\!=\!200$, $\Omega_{c}\!=\!2\Gamma_{e}'$. (a)-(d) we average over 1000 samples of atomic spatial distributions with $\mathcal {E}=0.0001\sqrt{\frac{\Gamma_{_{1D}}}{2c}}$, $\Gamma_{_{1D}}\!=\!2\Gamma_{e}'$, $k_{in}d\!=\!\pi/2$, and $\Delta_{c}\!=\!0$.} \label{figure3}
\end{figure}


In Figs. \ref{figure3}(a)-(b), we plot the transmission and reflection  of the incident field with detuning $\Delta/\Gamma_{e}'$ for
different values of the control beam Rabi frequency $\Omega_{c}$, averaged over 1000 samples of atomic spatial distributions.
First, we consider the case that the levels $|g\rangle$ and $|s\rangle$ are two hyperfine states in the ground state manifold, where the level $|s\rangle$ is metastable.
We observe that the atomic ensemble becomes fully transparent when the detuning is zero in the presence of the control field, which is known as EIT \cite{FleischhauerREV2005}. In fact, this phenomenon derives from destructive interference between two allowed atomic transitions, which causes the cancellation of the population of the excited state $|e\rangle$.
As shown in Figs. \ref{figure3}(c)-(d), we calculate the width $w$ of the central transparency window near two-photon resonance. Here, the width $w$ of the EIT window is defined by $T\!=\!T_{_{\Delta=0}}\text{exp}(-\Delta^{2}/w^{2})$ \cite{Lambropoulos}, which only holds for small $\Delta$.  We observe that the width of the EIT
window is proportional to the parameters $\Omega_{c}^2/\Gamma_{_{1D}}$ and $\frac{1}{\sqrt{n}}$, which agrees with the results of the one atom case \cite{FangPE2016} and linear array of superconducting artificial atoms \cite{Leung2012}.
While, different from single three-level atom case \cite{DRoyPRA2014,FangPE2016}, we see that the transmission is almost zero in two regions of the frequency detuning, and such a band-gap-like structure is the result of the scattering of multiple atoms.
In fact, by controlling the coupling strength $\Gamma_{_{1D}}$ and the number $n$ of atoms, we can tune the bandwidth.

\begin{figure}[tpb]    
\centering\includegraphics[width=8.4cm]{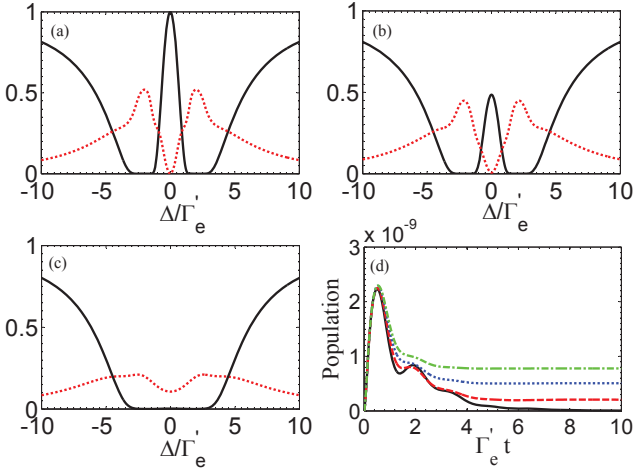}
\caption{The average transmission $\overline{T}$ (black solid line) and reflection $\overline{R}$ (red dotted line) spectra of the input field as a function of the frequency detuning $\Delta/\Gamma_{e}'$ for (a) $\Gamma_{t}'\!=\!0$, (b) $\Gamma_{t}'\!=\!0.3\Gamma_{e}'$, (c) $\Gamma_{t}'\!=\!3.5\Gamma_{e}'$. (d) The population of the collective atomic excitation $|E\rangle$ versus time for the total decoherence rate $\Gamma_{t}'\!=\!0$ (black solid line), $\Gamma_{t}'\!=\!0.5\Gamma_{e}'$ (red dashed line), $\Gamma_{t}'\!=\!1.0\Gamma_{e}'$ (blue dotted line), $\Gamma_{t}'\!=\!1.5\Gamma_{e}'$ (green dashed-dotted line). (a)-(d) n=10 atoms are randomly placed in a lattice of N=200 sites, and we average over 1000 samples of atomic spatial distributions and set the parameters $\mathcal {E}=0.0001\sqrt{\frac{\Gamma_{_{1D}}}{2c}}$, $\Gamma_{_{1D}}\!=\!2\Gamma_{e}'$, $\Omega_{c}\!=\!2\Gamma_{e}'$, $k_{in}d\!=\!\pi/2$, and $\Delta_{c}\!=\!0$. } \label{figure4}
\end{figure}

Since the influence of the decoherence in EIT-based light storage is important,  we also analyze the effect of decoherence originating from both population relaxation and dephasing between the two states $|g\rangle$ and $|s\rangle$. In actual atomic systems, the population relaxation between the two states $|g\rangle$ and $|s\rangle$ is usually caused by inelastic atom-atom and atom-wall collisions, and the dephasing of the forbidden $|g\rangle\!-\!|s\rangle$ transition exists due to elastic atom-atom and atom-wall collisions, trapping potential and laser fluctuations \cite{FleischhauerREV2005,WangJin2010}. In our system, we assume that the decoherence of all atoms is identical. The photon-mediated dipole-dipole interactions between atoms can be described by a master equation $\dot{\rho}=-i[H_{I},\rho]+\mathcal {L}\rho$ for the atomic density operator $\rho$, where
\begin{eqnarray}     \label{eqaadd1}       
H_{I}=\!\!&&-{{\sum\limits_{j}^n}}\big[\Delta S_{ee}^{j}+(\Delta-\Delta_{c}) S_{ss}^{j}+\Omega_{c}(S_{es}^{j}+h.c.)\big]\nonumber\\
&&+\frac{\Gamma_{_{1D}}}{2}{{\sum\limits_{j,k}^n}}sin({k_{in}|z_{j}-z_{k}|})S_{eg}^{j}S_{ge}^{k},
\end{eqnarray}
and
\begin{eqnarray}     \label{eqaadd2}       
\mathcal {L}\rho\!=\!\!\!\!\!\!&&\!-\frac{\Gamma_{_{1D}}}{2}{{\sum\limits_{j,k}^n}}cos({k_{in}|z_{j}\!-\!z_{k}|})(S_{eg}^{j}S_{ge}^{k}\rho+\rho S_{eg}^{j}S_{ge}^{k}\nonumber\\
&&-2S_{ge}^{k}\rho S_{eg}^{j})-\frac{\Gamma_{e}^{'}}{2}{{\sum\limits_{j}^n}}\big(\{S_{ee}^{j},\rho\}-2S_{ge}^{j}\rho S_{eg}^{j}\big)\nonumber\\
&&-\frac{\Gamma_{p}^{'}}{2}{{\sum\limits_{j}^n}}\big(\{S_{ss}^{j},\rho\}\!-\!2S_{gs}^{j}\rho S_{sg}^{j}\!+\!\{S_{gg}^{j},\rho\}\!-\!2S_{sg}^{j}\rho S_{gs}^{j}\big)\nonumber\\
&&-\frac{\Gamma_{d}^{'}}{2}{{\sum\limits_{j}^n}}\big(\{S_{ss}^{j},\rho\}\!-\!2S_{ss}^{j}\rho S_{ss}^{j}\!+\!\{S_{gg}^{j},\rho\}\!-\!2S_{gg}^{j}\rho S_{gg}^{j}\big).\nonumber\\
\end{eqnarray}
The third term of Eq. (\ref{eqaadd2}) accounts for population relaxation between the two states $|g\rangle$ and $|s\rangle$, and the fourth term of Eq. (\ref{eqaadd2}) describes the dephasing effect. For simplicity, we assume that the population relaxation rate from $|g\rangle\!\rightarrow\!|s\rangle$ is the same as the rate from $|s\rangle\!\rightarrow\!|g\rangle$ and is given by $\Gamma_{p}^{'}$, and the dephasing rate between the two states $|g\rangle$ and $|s\rangle$ is given by $\Gamma_{d}^{'}$.
Thus, we define the total decoherence rate as $\Gamma_{t}'=\Gamma_{p}'+\Gamma_{d}'$. With the above assumptions, using the master equation approach \cite{mastereqarxiv}, we calculate the influence of the total decoherence rate $\Gamma_{t}'$ on the transmission and reflection spectra of the driven $\Lambda$-type atomic ensemble.  As shown in Fig. \ref{figure4}, different values of $\Gamma_{t}'$ have remarkable effect on both transmission and reflection. In detail, we observe that only when the total decoherence rate $\Gamma_{t}'\!=\!0$, the atomic ensemble coupled to the 1D waveguide can be fully transparent on resonance, as shown in Fig. \ref{figure4}(a). Furthermore, with the increment of $\Gamma_{t}'$, the values of the peaks in transmitted spectrum  decrease, and finally the EIT transparency window disappears. Interestingly, with a low total decoherence rate $\Gamma_{t}'$, the reflection on resonance is always zero, while, when $\Gamma_{t}'$ is large enough, the reflection on resonance turns to be nonzero, as shown in Fig. \ref{figure4}(c). We also give the time evolution of the population in the collective atomic excitation $|E\rangle$, where four cases are considered, i.e., $\Gamma_{t}'\!=\!0$, $\Gamma_{t}'\!=\!0.5\Gamma_{e}'$, $\Gamma_{t}'\!=\!1.0\Gamma_{e}'$, $\Gamma_{t}'\!=\!1.5\Gamma_{e}'$, as shown in Fig. \ref{figure4}(d). Note that, the collective atomic excitation is $|E\rangle=(1/\sqrt{n})\sum_{j}|e_{j}\rangle$, where $|e_{j}\rangle$ denotes the presence of an excitation in the $j^{th}$ atom with all other atoms in the ground state. We observe that, all the plots of the population show an initial sharp peak and decrease
to a constant value after a time scale. For a fixed driving field $\Omega_{c}$, the time scale for the system to reach the steady state is reduced as the total decoherence rate $\Gamma_{t}'$ is increased. Moreover, the population of the collective atomic excitation increases with the decoherence rate $\Gamma_{t}'$. As mentioned above, in an ideal EIT condition, there is no population in the excited state $|e\rangle$ for every atom, which is the consequence of the dark state originating from the destructive interference between the atomic transitions $|g\rangle\!\leftrightarrow\!|e\rangle$ and $|e\rangle\!\leftrightarrow\!|s\rangle$.
In fact, the presence of the decoherence rate $\Gamma_{t}'$ drives the system out of the dark state, and
the EIT phenomenon is removed.



\begin{figure}[tpb]    
\centering\includegraphics[width=6.3cm]{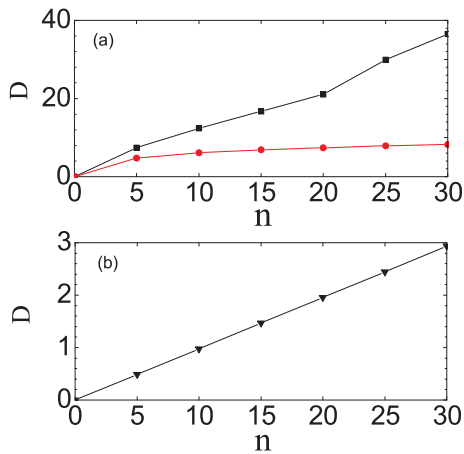}
\caption{ (a) The optical depth $D$ vs the number $n$ of atoms for $k_{in}d\!=\!\pi/2$ (black squares) and $k_{in}d\!=\!\pi$ (red dots) with $\Gamma_{_{1D}}\!=\!2\Gamma_{e}'$.
(b) The optical depth $D$ vs the number $n$ of atoms for $k_{in}d\!=\!\pi/2$ (black down triangles) with $\Gamma_{_{1D}}\!=\!0.05\Gamma_{e}'$.
Parameters: (a)-(b) $n$ atoms are randomly placed in a lattice of N=200 sites, and we average over $10^{6}$ samples of atomic spatial distributions and set the parameters $\mathcal {E}=0.0001\sqrt{\frac{\Gamma_{_{1D}}}{2c}}$, $\Omega_{c}\!=\!0$, $\Gamma_{t}'\!=\!0$, and $\Delta_{c}\!=\!0$. } \label{figque4}
\end{figure}

The field transmission in a medium is determined by the optical depth $D$, which is defined by $T(\Delta\!=\!0)= e^{-D}$ in the absence of the driving
field. As shown in Fig. \ref{figque4}(a), we calculate the optical depths for two choices of the lattice constant $d$ with a fixed coupling strength $\Gamma_{_{1D}}\!=\!2\Gamma_{e}'$. In detail, when $k_{in}d\!=\!\pi/2$, the optical depth $D$ increases quickly as we add the number of atoms, while for $k_{in}d\!=\!\pi$, the optical depth changes slowly with the number $n$ of atoms. Specifically, in the limit of $\Gamma_{_{1D}}\!\ll\!\Gamma_{e}'$, we find that the optical depth is given by
$D\approx2n\Gamma_{_{1D}}/\Gamma_{e}'$, as shown in Fig. \ref{figque4}(b). Since a medium requires a large optical depth for high storage efficiency in quantum memory \cite{Nilsson2005,Lvovsky2009}, in our system, we can obtain a requisite optical depth by controlling the number $n$ of atoms and the lattice constant $d$ with suitably large $\Gamma_{_{1D}}/\Gamma_{e}'$.

\begin{figure}[tpb]    
\centering\includegraphics[width=8.4cm]{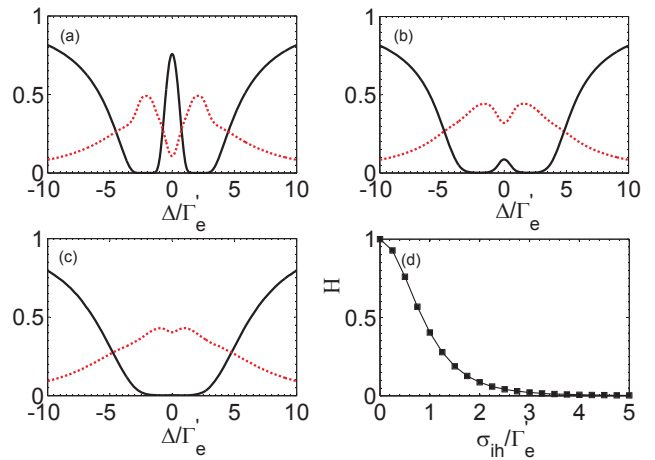}
\caption{ The average transmission $\overline{T}$ (black solid line) and reflection $\overline{R}$ (red dotted line) spectra of the input field as a function of the frequency detuning $\Delta/\Gamma_{e}'$ for (a) $\sigma_{ih}\!=\!0.5\Gamma_{e}'$, (b) $\sigma_{ih}\!=\!2\Gamma_{e}'$, (c) $\sigma_{ih}\!=\!5\Gamma_{e}'$. (d) The height $H$ of the EIT peak as a function of the parameter $\sigma_{ih}$ in the inhomogeneous broadening. (a)-(d) $n\!=\!10$ atoms are randomly placed in a lattice of $N\!=\!200$ sites, and we average over $10^{6}$ samples of atomic spatial distributions with $\mathcal {E}=0.0001\sqrt{\frac{\Gamma_{_{1D}}}{2c}}$, $\Gamma_{_{1D}}\!=\!2\Gamma_{e}'$, $k_{in}d\!=\!\pi/2$, $\Gamma_{t}'\!=\!0$, $\Omega_{c}\!=\!2\Gamma_{e}'$, and $\Delta_{c}\!=\!0$.  } \label{figque3}
\end{figure}

In the above calculations, we assume that all the atoms trapped in the lattice are identical with homogeneous broadening. While, in practice, the emitters in different lattice sites experience different trapping potentials, which affect the transition frequencies of the emitters according to their locations in the lattice. The broadening caused by such effect is inhomogeneous. In our system, the effect can probably happen for both the excited state and the metastable state. But since EIT only depends on the two-photon detuning, it is reasonable just to assume that the metastable state energy is shifted. In the following, we assume that the inhomogeneous broadening is Gaussian with the lineshape $g_{_{ih}}(\Delta_{ih})\!=\!\frac{1}{\sigma_{_{ih}}\sqrt{2\pi}}exp({-\frac{\Delta_{_{ih}}^{2}}{2\sigma_{_{ih}}^{2}}})$, where $2\sigma_{_{ih}}$ is the full width at half maximum of the lineshape in inhomogeneous broadening, and $\Delta_{ih}$ is the inhomogeneous detuning from the metastable level $|s\rangle$.


%

\begin{figure*}[tpb]    
\begin{center}
\centering\includegraphics[width=17.3cm]{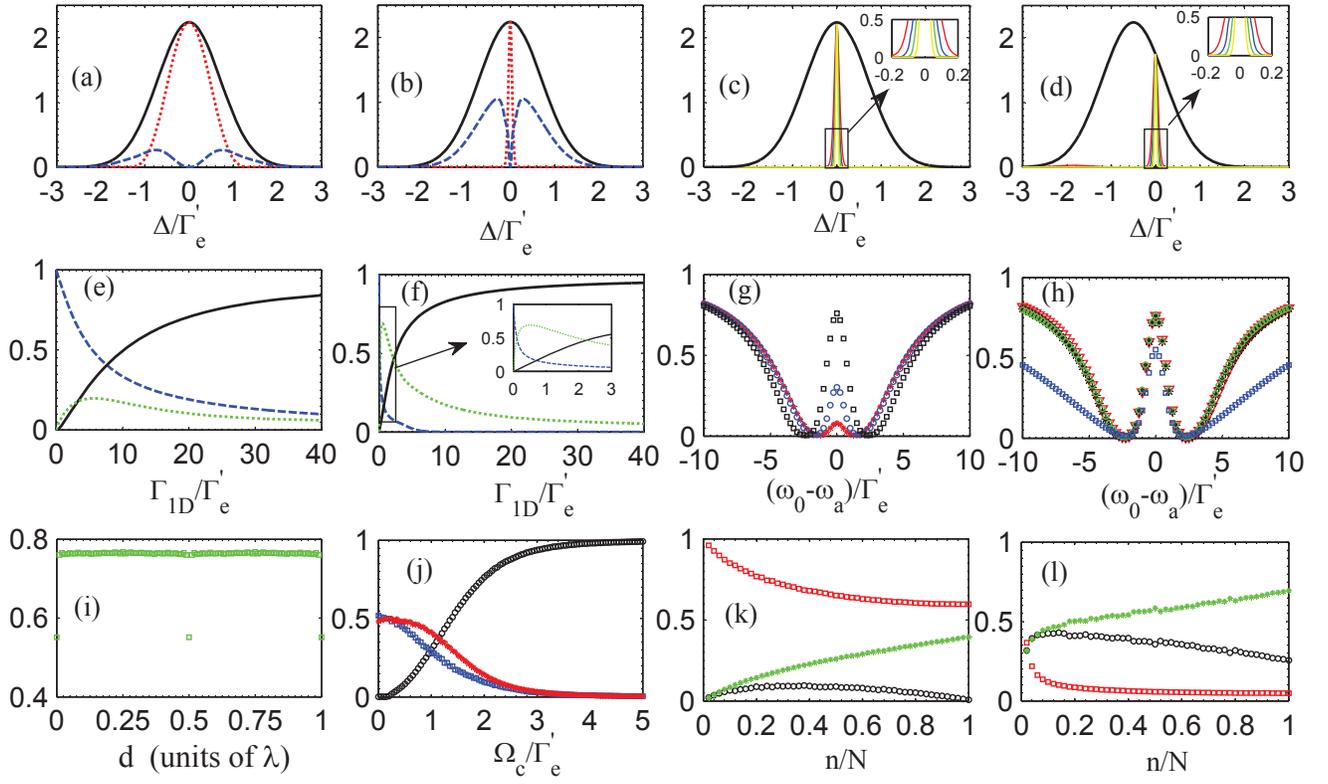}
\caption{ The spectra of the incident (black solid line), reflected (blue dashed line), and transmitted (red dotted line) photons with (a) $\Omega_{c}\!=\!2\Gamma_{e}'$, and (b) $\Omega_{c}\!=\!0.5\Gamma_{e}'$. The spectra of the incident (black), and transmitted photons for $n\!=\!5$ (red), 10 (blue), 20 (green), 50 (yellow) atoms randomly placed in a lattice of $N\!=\!200$ sites with  $\Omega_{c}\!=\!0.5\Gamma_{e}'$, (c) $\omega_{0}\!=\!\omega_{a}$, and (d) $\omega_{0}\!=\!\omega_{a}-0.5\Gamma_{e}'$.
The reflection (black squares), transmission (blue circles), and loss (green up triangles) as a function of the coupling strength $\Gamma_{_{1D}}/\Gamma_{e}'$ with (e) $\Omega_{c}\!=\!2\Gamma_{e}'$, and (f) $\Omega_{c}\!=\!0.5\Gamma_{e}'$. (g) The transmission as a function of center frequency deviation  for different driving fields $\Omega_{c}\!=\!0.5\Gamma_{e}'$ (red asterisks), $\Omega_{c}\!=\!1.0\Gamma_{e}'$ (blue circles), and $\Omega_{c}\!=\!2\Gamma_{e}'$ (black squares). (h)  The transmission as a function of center frequency deviation  for different lattice constant  $k_{in}d\!=\!\pi$ (blue squares), $k_{in}d\!=\!0.75\pi$ (green circles), $k_{in}d\!=\!0.5\pi$ (black asterisks), and $k_{in}d\!=\!0.25\pi$ (red down triangles) with $\Omega_{c}\!=\!2\Gamma_{e}'$. (i) The transmission as a function of lattice constant $d$ with $\Omega_{c}=2\Gamma_{e}'$. $\lambda$ is the wavelength resonant to the atomic transition $|g\rangle\!\leftrightarrow\!|e\rangle$. (j) The transmission (black circles), reflection (blue squares) and loss (red asterisks) as a function of the driving fields $\Omega_{c}/\Gamma_{e}'$. The transmission (red squares), reflection (black circles), and loss (green asterisks) as a function of the filling factor $n/N$ with (k) $\Omega_{c}\!=\!2\Gamma_{e}'$, and (l) $\Omega_{c}\!=\!0.5\Gamma_{e}'$. Parameters: (a)-(l) we average over 1000 samples of atomic spatial distributions with  $\Gamma_{t}'\!=\!0$, $\Delta_{c}\!=\!0$, $\sigma_{ih}\!=\!0$, and $\mathcal {E}=0.0001\sqrt{\frac{\Gamma_{_{1D}}}{2c}}$, (a)-(g) and (i)-(l) $k_{in}d\!=\!\pi/2$, (a)-(d) and (g)-(l) $\Gamma_{_{1D}}\!=\!2\Gamma_{e}'$, (a)-(b), (e)-(f), and (i)-(l) $\omega_{0}\!=\!\omega_{a}$, (a)-(b) and (e)-(j) $n\!=\! 10$ atoms are randomly placed in a lattice of $N\!=\!200$ sites, (k)-(l) the number of the sites in the lattice is $N\!=\!50$. } \label{figure5}
\end{center}
\end{figure*}

To proceed, we evaluate the effect of the inhomogeneous broadening on the transport properties of the incident field. As shown in Figs. \ref{figque3}(a)-(c), we give the transmission and reflection spectra of the input field in three cases, i.e., $\sigma_{_{ih}}\!=\!0.5\Gamma_{e}', 2\Gamma_{e}', 5\Gamma_{e}'$. We observe that, in contrast to the case with homogeneous broadening shown in Fig. \ref{figure3}(b), the inhomogeneous broadening of the emitters has remarkable influence on the transport properties of the input field. In detail,
for the transmission, the value of the EIT peak decreases quickly with the increment of $\sigma_{_{ih}}$. Interestingly, when $\sigma_{ih}\!>\!5.0\Gamma_{e}'$, the EIT phenomenon will almost completely disappear, i.e., $T(\Delta\!=\!0)\!\approx\!0$.
For the reflection, with the presence of the inhomogeneous broadening of the emitters, the value of the dip becomes nonzero, which means that the input field is partly reflected by the emitters at $\Delta\!=\!0$. Moreover, we study the height of the EIT peak as
a function of the parameter $\sigma_{_{ih}}$ in the inhomogeneous broadening, as shown in Fig. \ref{figque3}(d). We observe that, when we increase the parameter $\sigma_{_{ih}}$, the height of the EIT peak decreases. In other words, the EIT phenomenon is sensitive to the parameter $\sigma_{_{ih}}$ in the inhomogeneous broadening of the metastable level $|s\rangle$.

\subsection{ An input field with Gaussian shape}   \label{Gaussian}

In practice, the input field is a pulse with finite bandwidth. Here, we study the scattering property of a Gaussian pulse interacting with the atomic ensemble coupled to the 1D waveguide. In experiment, using a single-photon electric-optic modulation \cite{EOM}, we can produce a photon pulse with Gaussian shape given by
\begin{eqnarray}     \label{eqa12}       
A(\omega)=\frac{(8\pi)^\frac{1}{4}}{\sqrt{\sigma L}}e^{-(\omega-\omega_{0})^2/\sigma^{2}},
\end{eqnarray}
where $\sigma$ is the width in the frequency space with the full width at half maximum of the spectrum, $L$ is the quantization length in the propagation direction, and $\omega_{0}$ is the center frequency of the pulse.
$A(\omega)$ denotes the probability amplitude of the photon component at frequency $\omega$. Note that $(L/2\pi)\int_{-\infty}^{\infty}|A(\omega)|^{2}d\omega\!=\!1$ is the requirement for a single-photon number, and we set $\sigma\!=\!L\!=\!1$ in the following section.

As shown in Fig. \ref{figure5}(a), we calculate the spectra of the incoming, reflected, and transmitted fields with $\omega_{0}\!=\!\omega_{a}$, $\Omega_{c}\!=\!2\Gamma_{e}'$, and $\Gamma_{_{1D}}\!=\!2\Gamma_{e}'$. We observe that the spectrum of the transmitted photon is similar to the initial shape of the incident photon and the photon component around the atomic resonance frequency $\omega_{a}$ can transmit the atomic ensemble completely, which is the result of EIT shown in Fig. \ref{figure3}(a). While, the spectrum of the reflected component is different and has two peaks, which originates from the two peaks in the reflection spectrum shown in Fig. \ref{figure3}(a). However, when we turn the condition $\Omega_{c}\!=\!2\Gamma_{e}'$ to $\Omega_{c}\!=\!0.5\Gamma_{e}'$ with other parameters remaining unchanged, we can get some different results shown in Fig. \ref{figure5}(b): compared with the case of $\Omega_{c}\!=\!2\Gamma_{e}'$, the spectrum of the transmitted photon becomes narrower and the values of the peaks in the spectrum of the reflected part turn larger when $\Omega_{c}\!=\!0.5\Gamma_{e}'$. This is because, when we decrease the Rabi frequency $\Omega_{c}$, the width of the EIT window will be reduced, and the splitting
of the two peaks in the reflection spectrum decreases, as shown in Fig. \ref{figure3}. With more calculations, we conclude that when $\frac{\Omega_{c}}{\Gamma_{e}'}\gg1$, the shape of the transmitted photon is very similar to the input photon. While when $\frac{\Omega_{c}}{\Gamma_{e}'}\ll1$, a Lorentzian peak appears at the frequency $\omega\!=\!\omega_{a}$ in the spectrum of the transmitted pulse. That is, the transmitted spectrum of the Gaussian pulse can be effectively controlled by tuning the Rabi frequency $\Omega_{c}$ of the driving field.

The spectra of the transmitted fields with different numbers of atoms under the condition $\omega_{0}\!=\!\omega_{a}$, $\Omega_{c}\!=\!0.5\Gamma_{e}'$, and $\Gamma_{_{1D}}\!=\!2\Gamma_{e}'$ are shown in Fig. \ref{figure5}(c). Here, four cases are considered: $n\!=\!5$ (red), 10 (blue), 20 (green), 50 (yellow) atoms are randomly placed in a lattice of $N\!=\!200$ sites, and we average over 1000 samples of atomic spatial distributions for every case. We see that, when more atoms are placed in the lattice, the Lorentzian peak in the spectrum of the transmitted field becomes narrower. This is because the width of the EIT window will decrease when more atoms are placed in the system, as shown in Fig. \ref{figure3}(d). Moreover, we study a more general case where the center frequency $\omega_{0}$ of the incident Gaussian pulse is different from atomic resonance frequency $\omega_{a}$. For example, the transmitted spectrum under the condition $\omega_{0}\!=\!\omega_{a}-0.5\Gamma_{e}'$ is shown in Fig. \ref{figure5}(d).  Although $\omega_{0}\neq\omega_{a}$, the number of atoms has the same effect on the spectrum, i.e., the component of the incident field at the resonance frequency can transmit the atomic ensemble completely. As shown in Figs. \ref{figure5}(a)-(d), by tuning Rabi frequency of the driving field and the number of atoms, we can only transmit the frequency component $\omega_{a}$ of the Gaussian pulse completely, and the other parts of the pulse will be reflected or decay into the free space. That is, our system may be useful as a photon frequency filter, which circumvents the challenge of integrating the waveguide system with other optical components.

The reflection, transmission, and loss as a function of coupling strength under the condition $\Omega_{c}\!=\!2\Gamma_{e}'$ are shown in Fig. \ref{figure5}(e). We  see that the transmission (reflection) of the Gaussian pulse decreases (increases) when we increase the coupling strength $\Gamma_{_{1D}}$, while the loss first increases and then decreases to a constant value (not zero) as we enhance the coupling strength. When
$\Gamma_{_{1D}}\!\approx\!5.75{\Gamma_{e}'}$, the loss of the incident photon pulse reaches the maximum value 19.7\%. However, as we only change the condition $\Omega_{c}\!=\!2\Gamma_{e}'$ to be $\Omega_{c}\!=\!0.5\Gamma_{e}'$, the results are different, as shown in Fig. \ref{figure5}(f).
We observe that the variation trends of the reflection, transmission, and loss with the coupling strength are the same, while they all change more rapidly than the results with $\Omega_{c}\!=\!2\Gamma_{e}'$. Moreover, the transmission will approach zero when the coupling strength is large enough in both cases. While, it is not easy to obtain strong coupling between the atomic ensemble and the 1D waveguide in experiment, and Figs. \ref{figure5}(e)-(f) show us that the transport properties of the system can be controlled by tuning the Rabi frequency $\Omega_{c}$ of the driving field, which should be more convenient.

We also study the transmission as a function of the detuning between the center frequency $\omega_{0}$ of the incident Gaussian pulse and atomic resonance frequency $\omega_{a}$. The results are shown in Fig. \ref{figure5}(g). Here, we consider three choices of the driving fields, i.e., $\Omega_{c}\!=\!0.5\Gamma_{e}', 1\Gamma_{e}', 2\Gamma_{e}'$. The similarities are: (1) a peak appears at the frequency $\omega_{0}\!=\!\omega_{a}$ in the transmitted spectrum, which is actually the result of EIT, (2) two dips exist when the center frequency of the Gaussian pulse is red and blue detuned from the atomic resonance frequency, (3) the incident photon pulse will transmit the atomic ensemble with no interaction when ($\omega_{0}-\omega_{a})\gg\Gamma_{e}'$. However, with different choices of the driving fields, the values of the peaks at the frequency $\omega_{0}\!=\!\omega_{a}$ are quite different. When the Rabi frequency of the driving field is $\Omega_{c}\!=\!2\Gamma_{e}'$, the value of the peak can be 75.9\%. While, when it is changed to be $\Omega_{c}\!\!=\!\!1\Gamma_{e}'$ ($0.5\Gamma_{e}'$), the value of the peak drops down to 30.2\% (8.2\%). This is because the Rabi frequency $\Omega_{c}$ of the driving field influences the width of the EIT window, as shown in Fig. \ref{figure3}(c). Therefore, to effectively control the transmission of the incident Gaussian pulse, one way is changing the
Rabi frequency of the driving field. The other way is changing the center frequency $\omega_{0}$ of the Gaussian pulse. Furthermore, the transmission as a function of the detuning $(\omega_{0}-\omega_{a})$ for different choices of $k_{in}d$ is shown Fig. \ref{figure5}(h). When $k_{in}d\!=\!0.25 \pi$, $k_{in}d\!=\!0.5 \pi$, and $k_{in}d\!=\!0.75 \pi$, the shapes of functions are very similar. However, with $k_{in}d\!=\!\pi$, for almost the whole region of the detuning $(\omega_{0}-\omega_{a})$, the transmission of the Gaussian pulse becomes smaller than those in the three cases mentioned above. To show clearly the influence of lattice constant $d$ on the transmission of the Gaussian pulse, we plot Fig. \ref{figure5}(i) with $\Omega_{c}\!=\!2\Gamma_{e}'$. An obvious difference appears in the transmission when the lattice constant is $d\!=\!0, \lambda/2, \lambda$, respectively. While, for any other choices of $d$, the values of the transmission are basically the same. In fact, this phenomenon is caused by the last part of the Hamiltonian in Eq. (\ref{eqa9}). In the three special cases $d\!=\!0, \lambda/2, \lambda$, for any possible configurations of atomic positions, the imaginary component of $e^{ik_{in}|z_{j}-z_{k}|}$, i.e., $i\sin(k_{in}|z_{j}-z_{k}|)$ is always zero, which changes the transmission of the incident pulse dramatically.
Moreover, we give the transmission, reflection, and loss as a function of the driving field $\Omega_{c}$, as shown in Fig. \ref{figure5}(j). We observe that, as we enhance the driving field, the transmission increases from zero rapidly, and inversely, both the reflection and loss decrease to zero quickly. When the Rabi frequency of the driving field is large enough, for example, $\Omega_{c}\!=\!3.5\Gamma_{e}'$, the transmission will approach 100\%, and both the reflection and loss will touch zero. That is, by changing the driving field, we can effectively control the transport properties of the incident photon pulse, which is consistent with the results shown in
Fig. \ref{figure5}(g).

\begin{figure} 
\centering\includegraphics[width=8.3cm]{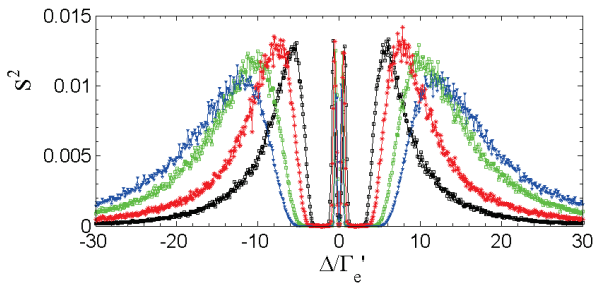}
\caption{  The variance $s^{2}$ of the transmission $T$ when $n\!=\!10$ (black line with squares), $n\!=\!20$ (red line with asterisks), $n\!=\!40$ (green line with circles), $n\!=\!60$ (blue line with down triangles) atoms are placed randomly over $N\!=\!200$ sites.  Here, 1000 samples of atomic spatial distributions are averaged per detuning with  $k_{in}d=\pi/2$, $\mathcal {E}=0.0001\sqrt{\frac{\Gamma_{_{1D}}}{2c}}$, $\Omega_{c}\!=\!2\Gamma_{e}'$, $\Gamma_{t}'\!=\!0$, $\sigma_{ih}\!=\!0$, and $\Delta_{c}\!=\!0$.  } \label{figure6}
\end{figure}

Finally, we study the transmission, reflection, and loss as a function of the filling factor $n/N$ for two different choices of the driving fields with $N\!=\!50$. As shown in Fig. \ref{figure5}(k), when $\Omega_{c}\!=\!2\Gamma_{e}'$, the transmission decreases slowly from 1 to a nonzero value as we add the number of atoms, while for the reflection, it first increases from zero slowly and then decreases to zero slowly. Moreover, when the filling factor is small ($0\!<\!n/N\!\leq\!0.2$), the loss scales nonlinearly with the filling factor, when the filling factor is large ($0.2\!<\!n/N\!\leq\!1.0$), the loss scales linearly with the filling factor. For $\Omega_{c}\!=\!0.5\Gamma_{e}'$ with other parameters remaining unchanged, the results are shown in Fig. \ref{figure5}(l). Similarly, the variation trends of the reflection, transmission, and loss with the filling factor are basically the same. Differently, we observe that, with the equal number of the atoms, both the loss and the reflection of the Gaussian pulse in this case are larger than that for $\Omega_{c}\!=\!2\Gamma_{e}'$, while the transmission in this case becomes much smaller than that for $\Omega_{c}\!=\!2\Gamma_{e}'$. In other words, the driving field influences the decay rate of the atoms out of the waveguide, i.e., the stronger the driving field, the weaker the loss, which is consistent with the results shown in Fig. \ref{figure5}(j).

\subsection{Transmission variance caused by atomic spatial distributions}  \label{variance}

Different from the previous work where the atoms are equally located  with a deterministic separation, we focus on the case that the atoms are randomly placed in a lattice along the waveguide. In our scheme, due to the various configurations of atomic positions, the scattering properties of the incident field are variational. Here, to describe the influence on the transmission caused by atomic spatial distributions, we use the variance $s^{2}$, which is defined as
\begin{eqnarray}     \label{eqa13}       
s^{2}=\frac{1}{m}\sum\limits_{i}^m(T_{i}-\bar{T})^2,
\end{eqnarray}
where $m$ denotes the sample size of atomic spatial distributions, $T_{i}$ is the transmission for the $i^{th}$ sample, and $\bar{T}$ is the average transmission for all samples.

As shown in Fig. \ref{figure6}, we obtain the variance $s^{2}$ of the transmission as a function of the detuning for 1000 samples when $n\!=\!10$ atoms are randomly placed in $N\!\!=\!\!200$ sites. We observe that the plot is symmetric, and $s^{2}$ is zero in a range of the frequency detuning around $\Delta\!=\!\pm\Omega_{c}$ and when $\Delta\!=\!0$, which is the result of EIT and the band-gap-like structure in transmission spectrum shown in Figs. \ref{figure3}(a)-(b). There are two peaks around the detuning $\Delta\!=\!\Omega_{c}$ ($\Delta\!=\!-\Omega_{c}$), i.e., when the detuning is shifted around $\Delta\!\!=\!\!\pm\Omega_{c}$, the influence of atomic spatial distributions on transmission become obvious. Moreover, $s^{2}$ will approach zero for a large detuning, which corresponds to the case that the incident field transmits the atomic ensemble with no interaction, and the transmission is not affected by atomic spatial distributions. We also study the variance $s^{2}$ of the transmission $T$ for different choices of the number of atoms. Here, we consider another three cases, i.e., the number of atoms is $n \!\!=\!\!20, 40, 60$, respectively. We see that the width of the dip near the Rabi frequency of the driving field is determined by the number of atoms when the sites of lattice $N$ is fixed. In detail, as we add the number of atoms, the width of the dip around the Rabi frequency of the driving field increases.
Moreover, for the region of the detuning $12\!\leq\!|\Delta|/\Gamma_{e}'\!\leq\!30$, when we increase the number of atoms, the value of the variance $s^{2}$ becomes larger. The results show that more atoms bring more fluctuation on the transmitted spectrum for a fixed sites $N$ of the lattice.

\subsection{Two-photon correlation}  \label{correlation}

\begin{figure*} [tpb]    
\begin{center}
\centering\includegraphics[width=14cm]{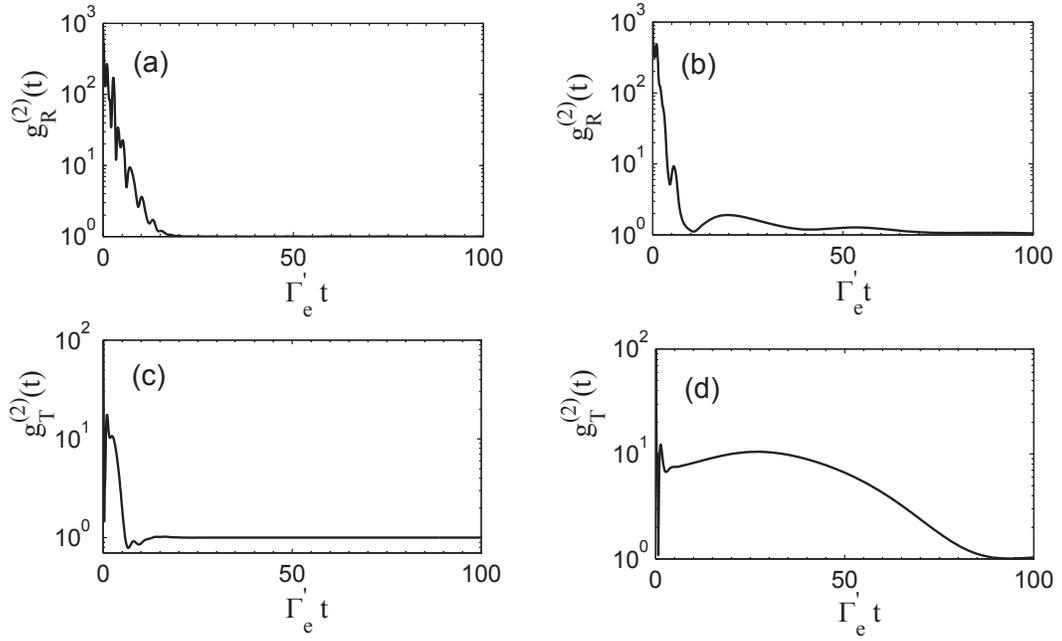}
\caption{ The photon-photon correlation function $\text{g}^{(2)}(t)$ of the output field when $n\!=\!10$ atoms are placed randomly over $N\!=\!200$ sites. The first row is for transmitted field with the driving field (a) $\Omega_{c}\!=\!2\Gamma_{e}'$, and (b) $\Omega_{c}\!=\!0.5\Gamma_{e}'$.
The second row is for reflected field with the driving field (c) $\Omega_{c}\!=\!2\Gamma_{e}'$, and (d) $\Omega_{c}\!=\!0.5\Gamma_{e}'$.
Here, the frequencies of the incident photons are chosen as one of the frequencies for $T\!=\!R$, and we average 1000 samples of atomic spatial distributions with $\mathcal {E}=0.0001\sqrt{\frac{\Gamma_{_{1D}}}{2c}}$, $k_{in}d\!=\!\pi/2$, $\Delta_{c}\!=\!0$, $\sigma_{ih}\!=\!0$, and $\Gamma_{t}'\!=\!0$.
} \label{figure7}
\end{center}
\end{figure*}

The main signature of non-classical light is that the photons can be bunched or anti-bunched, which can be calculated by photon-photon correlation function $\text{g}^{(2)}$ (also called the second-order coherence \cite{Loudon2003}). The two-photon correlation functions for two-level and three-level atoms coupled to an infinite waveguide have been considered in the  previous works \cite{ZhengPRL2013,LaaksoPRL2014,FangPRA2015,FangPE2016}. For a steady state, $\text{g}^{(2)}$ of the output field is defined as
\begin{eqnarray}     \label{eqa14}       
\text{g}^{(2)}(\tau)\!\!=\!\!\lim_{t\rightarrow\infty}\frac{\langle a^{\dagger}(z,t)a^{\dagger}(z,t+\tau)a(z,t+\tau)a(z,t)\rangle}{\langle a^{\dagger}(z,t)a(z,t)\rangle\langle a^{\dagger}(z,t+\tau)a(z,t+\tau)\rangle}.\nonumber\\
\end{eqnarray}
In our system, we can switch this definition to the Schr$\ddot{o}$dinger picture:
\begin{eqnarray}     \label{eqa15}       
\text{g}_{\alpha}^{(2)}(\tau)\!\!=\!\!\frac{\langle\psi|a_{_{\alpha}}^{\dagger}(z)e^{iH\tau}a_{_{\alpha}}^{\dagger}(z)a_{_{\alpha}}(z)e^{-iH\tau}a_{_{\alpha}}(z)|\psi\rangle}{|\langle\psi|a_{_{\alpha}}^{\dagger}(z)a_{_{\alpha}}(z)|\psi\rangle|^{2}},
\end{eqnarray}
where $|\psi\rangle$ is the steady-state wavevector, and $\alpha\!=\!R, T$.

Now, with a weak probe field ($\sqrt{\frac{c\Gamma_{_{1D}}}{2}}\mathcal {E}\!\!\ll\!\!\Gamma_{e}^{'}$), we discuss photon-photon correlation function $\text{g}^{(2)}$ for two choices of the driving fields in the off-resonant case when $n\!=\!10$ three-level atoms are randomly placed over $N\!=\!200$ sites. Here, the frequencies of the two identical photons are chosen as one of the frequencies for $T\!\!=\!\!R$, which are labeled by blue lines in Figs. \ref{figure3}(a)-(b). As shown in Fig. \ref{figure7}, we observe that strong initial bunching ($\text{g}^{(2)}\!\!>\!\!1$) is present for both reflection and transmission. Differently, as the time $\Gamma_{e}'t$ increases,  bunching dominates at the whole time scale with quantum beats (oscillation) for reflection $\text{g}_{_{R}}^{(2)}$ with both $\Omega_{c}\!=\!2\Gamma_{e}'$ and $\Omega_{c}\!=\!0.5\Gamma_{e}'$.
While for transmission $\text{g}_{_{T}}^{(2)}$, when $\Omega_{c}\!=\!2\Gamma_{e}'$, the initial bunching is followed by anti-bunching ($\text{g}^{(2)}\!\!<\!\!1$) with a small region of the parameter $\Gamma_{e}'t$, as shown in Fig. \ref{figure7}(c). When $\Omega_{c}\!=\!0.5\Gamma_{e}'$, no anti-bunching appears in the reflected field, which is shown in Fig. \ref{figure7}(d). Moreover, the intensity of the driving field $\Omega_{c}$ has an obvious influence on the correlations properties. By comparing the first and second columns of Fig. \ref{figure7}, we find that, for both the transmitted and reflected fields, when we enhance the driving field, the timescale for the decay of the two-photon correlations will be considerably shortened with more oscillations.

Specifically, on resonance $\Delta\!=\!0$, since the incident photons can transmit the atomic ensemble with $100\%$, no correlation will be generated. The correlation function of the transmitted field is $\text{g}_{_{T}}^{(2)}\!=\!1$ (not shown), which is consistent with the results in Refs. \cite{DRoyPRA2014,ZhengHXpra2012}. Actually, this phenomenon is known as ``fluorescence quenching" \cite{ZhouPRL1996,RephaeliPRa2011} and is not influenced by the parameters of our system, such as the number of atoms, the driving field, lattice constant $d$, and atomic spatial distributions. The above calculations show that our system  may provide an effective candidate for producing non-classical light in experiment.

\section{Conclusion}   \label{discussion}

In summary, with an effective non-Hermitian Hamiltonian, we have explored the interaction between a weak input field and an ensemble of $\Lambda$-type
three-level atoms coupled to a 1D waveguide. In our system, the atoms are randomly located in the lattice along the axis of the 1D waveguide, and we calculate the statistical properties by adopting the average values from a large sample of atomic spatial distributions. EIT is observed for the driven $\Lambda$-type atomic ensemble coupled to the waveguide, and the width of the EIT window is proportional to the parameters $\Omega_{c}^2/\Gamma_{_{1D}}$ and $\frac{1}{\sqrt{n}}$. We calculate the influence of decoherence on the transmission and reflection spectra of the driven $\Lambda$-type atomic ensemble. We conclude that, to maintain the EIT phenomenon, $\Gamma_{t}'$ must be much smaller than the coupling strength $\Gamma_{_{1D}}$. Moreover, we analyze the effect of the inhomogeneous broadening on the transmission and reflection spectra of the incident field, and find that the EIT phenomenon is very sensitive to the parameter $\sigma_{_{ih}}$ in the inhomogeneous broadening of the metastable level $|s\rangle$. Then, we adopt a pulse with Gaussian shape as the incident field, and analyze the rich optical properties with the parameters. The results show that, we can effectively control the transport properties of the input pulse by tuning the Rabi frequency of the driving field, the number of atoms, and the lattice constant $d$. Besides, by calculating the variance of the transmission caused by atomic spatial distributions, we find that the variance can approach zero in some region of the frequency detuning, which indicates that the transmission of the incident pulse is not affected by atomic spatial distributions. Moreover, we calculate the photon-photon correlation of the output fields generated by the scattering between the incident field and the atomic ensemble coupled to the 1D waveguide, which shows non-classical behavior such as bunching and anti-bunching. That is, the scattering between an input field and atomic ensemble in a 1D waveguide may provide an effective method for generating non-classical light in experiment.

\section*{ACKNOWLEDGMENTS}
GZS, FGD and GJY are supported by the National Natural Science
Foundation of China under Grants No. 11474026 and No.
11674033, and the Fundamental Research Funds for the
Central Universities under Grant No. 2015KJJCA01.
EM, WN and LCK acknowledge support from the National Research
Foundation and Ministry of Education, Singapore.

\end{document}